\documentclass[trackchanges,twocolumn,twocolappendix]{aastex701}

\begin{document}

\title{The Stellar-Mass Dependence of Feedback-Driven Outflows in DESI DR1 Star-Forming Galaxies}

\correspondingauthor{Yimeng Tang, Yifei Luo, Conghao Zhou}
\email{ymtang@ucsc.edu, yifeiluo@lbl.gov, zhou.conghao@ucsc.edu}

\author[0000-0003-2876-577X]{Yimeng Tang}
\affiliation{Department of Astronomy \& Astrophysics, University of California, Santa Cruz, 1156 High Street, Santa Cruz, CA 95064, USA}
\email{ymtang@ucsc.edu}

\author[0000-0001-7729-6629]{Yifei Luo}
\affiliation{Lawrence Berkeley National Laboratory, 1 Cyclotron Road, Berkeley, CA 94720, USA}
\email{yifeiLuo@lbl.gov}

\author[0000-0002-2897-6326]{Conghao Zhou}
\affiliation{Department of Physics, University of California, Santa Cruz, 1156 High Street, Santa Cruz, CA 95064 USA}
\email{zhou.conghao@ucsc.edu}

\begin{abstract}

Stellar feedback plays a key role in regulating star formation and shaping the evolution of star-forming galaxies. Low-mass galaxies provide one of the most powerful laboratories for testing stellar feedback and its impact on galaxy evolution. Numerical simulations predict that repeated feedback-driven gas inflows and outflows generate a ``breathing mode,'' producing a positive correlation between mass-normalized gas velocity dispersion and recent star formation in dwarf galaxies. Whether this kinematic signature persists at higher stellar masses has remained untested. We investigate this question using $\sim1200$ star-forming galaxies with Dark Energy Spectroscopic Instrument (DESI) spectroscopy and COSMOS2020 photometry spanning $7.9<\log M_\star/M_\odot<11.5$ within $z<0.25$. Ionized-gas kinematics are measured with {\tt pPXF}, and recent star formation histories are reconstructed through joint spectrophotometric fitting with {\tt Prospector}. We confirm the predicted positive correlation throughout the low-mass regime with $\log M_\star/M_\odot \lesssim 9.8$, quantitatively consistent with previous observations and FIRE-2 simulations. Above $\log M_\star/M_\odot \sim 9.8$, the correlation weakens rapidly and becomes consistent with zero, providing the first empirical constraint on the stellar mass scale above which the kinematic signature of feedback-driven outflow cycles becomes negligible. Our results establish a characteristic stellar mass above which stellar feedback no longer dominates galaxy-scale gas kinematics, providing a new observational benchmark for models of feedback-regulated galaxy evolution. Future joint analyses of gas kinematics and stellar structure will test whether this transition is accompanied by the broader galaxy response predicted by the breathing mode scenario.

\end{abstract}


\section{Introduction} 
\label{sec:intro}

Stellar feedback in star-forming galaxies is a central ingredient of galaxy evolution, regulating star formation, driving multiphase outflows, and redistributing and recycling gas within and around galaxies \citep{Somerville2015,Naab2017}. Its dynamical influence is expected to depend strongly on the mass because the ability of feedback to accelerate or expel gas decreases as gravitational potentials deepen \citep{Thompson2024,Collins2022}.

Low-mass ``dwarf'' galaxies ($\log M_\star/M_\odot \lesssim 9.5$) provide one of the sharpest observational tests of the small-scale challenges to the $\Lambda$CDM paradigm \citep[e.g.,][]{Weinberg2015,Bullock2017,Sales2022}. Their rotation curves often favor nearly constant-density dark matter (DM) ``cores'' \citep{Moore1994,Flores1994,McGaugh2001,deBlok2008,Oh2011,Oh2015}, in contrast to the steep central ``cusps'' predicted by dissipationless CDM-only simulations \citep{Navarro1996,Navarro2010,Springel2008}. A leading baryonic explanation invokes stellar feedback: repeated episodes of star formation drive gas outflows that temporarily reduce the central gravitational potential, causing non-adiabatic heating of DM and stellar orbits. As the gas subsequently cools and re-accretes, this process repeats, and over many cycles the central DM density is progressively reduced \citep{Governato2010,Pontzen2012}, producing DM cores.

A key prediction of this feedback-driven scenario is that baryons should respond to the same potential fluctuations that reshape the DM distribution. Using the FIRE-2 zoom-in simulations \citep{Hopkins2018}, \cite{ElBadry2016,ElBadry2017} showed that repeated gas inflow and outflow cycles drive a ``breathing mode'', in which phases of a deep gravitational potential, compact galaxy size, and high specific star formation rate (sSFR) alternate with phases of a shallow potential, expanded size, and low sSFR on timescales of a few hundred Myr, persisting to $z\sim0$. Similar bursty star formation cycles have been found in other simulations with different numerical methods and feedback prescriptions \citep[e.g.,][]{Sparre2017}.

Some ingredients of this picture are supported by observations. Bursty star formation is indicated by differences between H$\alpha$- and UV-based star formation rates (SFRs) in nearby dwarfs and at intermediate redshift \citep{Weisz2012,Guo2016,Emami2019}, while outflows are observed across the dwarf population with efficiencies increasing toward lower masses \citep{Chisholm2015,McQuinn2019}, consistent with the mass dependence predicted by simulations \citep{Muratov2015}.

These feedback-driven outflow and re-accretion cycles have a direct kinematic consequence. During burst phases, stellar feedback simultaneously enhances the ionized-gas velocity dispersion and recent star formation activity, whereas both quantities decline after the gas is expelled \citep{ElBadry2016,ElBadry2017}. The simulations therefore predict a positive correlation between mass-normalized gas velocity dispersion and sSFR \citep{Luo2026}, providing a direct observable signature of the breathing mode. Because the efficiency of this mechanism is determined by the competition between stellar feedback and the depth of the gravitational potential, it is expected to be strongest at $\log M_\star/M_\odot \sim$ 7--9 and to weaken progressively toward higher stellar masses as potential wells become deeper \citep[e.g.,][]{Governato2012,Pontzen2012,Tollet2016}. However, this predicted mass dependence has not yet been systematically tested in observation.

Several observational studies have explored such kinematic signatures. Using adaptive-optics spectroscopy of 17 gravitationally lensed dwarf galaxies at $z\sim2$, \citet{Hirtenstein2019} found broad agreement with the gas velocity dispersions predicted by FIRE-2, although their sample probes a higher-redshift and higher-SFR population than the nearby dwarfs that motivated the core--cusp problem. Other observational diagnostics, including SFR-driven size fluctuations \citep{Emami2021}, the anti-correlation between SFR and rotational support \citep{Wang2024}, and correlations between sSFR and gas velocity dispersion in larger spectroscopic samples \citep{Cicone2016,Pelliccia2020}, have also suggested a connection between recent star formation and galaxy dynamics. However, none of these studies were designed for the low-redshift dwarf regime with sufficient spectral resolution.

Most recently, \citet{Luo2026} performed the first quantitative test of this slope prediction between normalized gas velocity dispersion and sSFR, combining high resolution spectroscopy from Keck/DEIMOS with multi-band photometry for 103 star-forming dwarf galaxies ($7.9<\log M_\star/M_\odot<9.6$, $0.02<z<0.19$). Applying the same measurement to FIRE-2 galaxies, they found predicted slopes of $0.092\pm0.005$ for the average sSFR within 100 Myr and $0.115\pm0.004$ for 10 Myr, agreeing with $0.065\pm0.012$ and $0.096\pm0.012$, respectively, in their observations. This established the slope of this relation as a quantitative diagnostic of the breathing mode. However, their sample covered only a narrow stellar mass range in the dwarf regime, leaving the transition from feedback-dominated to gravitationally dominated gas kinematics unexplored. Because the amplitude of feedback-driven potential fluctuations should decline as galaxies become more massive, the key unresolved question is where the breathing mode signature weakens or disappears. Answering this question requires a sample spanning a substantially broader stellar mass range, together with star formation histories (SFHs) flexible enough to capture bursty activity. 

The Dark Energy Spectroscopic Instrument \citep[DESI;][]{DESI2016,DESI2022} is enabling statistical studies of dwarf galaxies through its wide area, highly multiplexed spectroscopy. Recent DESI studies have probed dwarf galaxies from both baryonic and dark matter halo perspectives. Combining DESI spectroscopy with ALFALFA observations, \citet{Scholte2024} identified changes in the atomic gas sequence and mass-metallicity relation near $M_\star\sim10^9\,M_\odot$. More recently, \citet{Shao2026} used DESI DR1 to extend direct constraints on the stellar-to-halo mass relation into the dwarf regime, further demonstrating the power of DESI to study mass-dependent galaxy evolution.

In this Letter, we address these limitations by combining DESI DR1 spectroscopy \citep{DESI2026} with COSMOS2020 photometry \citep{Weaver2022} for $\sim1200$ star-forming galaxies spanning nearly four decades in stellar mass. We reconstruct non-parametric SFHs jointly from the spectroscopic and photometric data, enabling a more direct probe of feedback-driven SFR variability on short timescales. All these data and methods are described in Section \ref{sec:sample}, and we present and discuss our results in Sections \ref{sec:disc_bursty}, \ref{sec:disc_transition} and \ref{sec:disc_cores}, followed by the conclusion in Section \ref{sec:summary}. A flat $\Lambda$CDM cosmology is adopted in the work, with $H_0=70\ {\rm km\ s^{-1}\ Mpc^{-1}}$, $\Omega_{\rm m}=0.3$, and $\Omega_{\rm \Lambda}=0.7$.

\section{Data and Method} 
\label{sec:sample}

Our analysis requires two measurements for each galaxy: the ionized-gas velocity dispersion and the recent sSFR. We describe the data and sample selection in Section \ref{sec:sample_description}, the ionized-gas kinematics measurements in Section \ref{sec:velocity_dispersion}, and the SFH reconstruction from joint spectrophotometric fitting in Section \ref{sec:prospector}.

\subsection{The DESI$\times$COSMOS2020 Sample}
\label{sec:sample_description}

\begin{figure}
\centering
\includegraphics[width=\columnwidth]{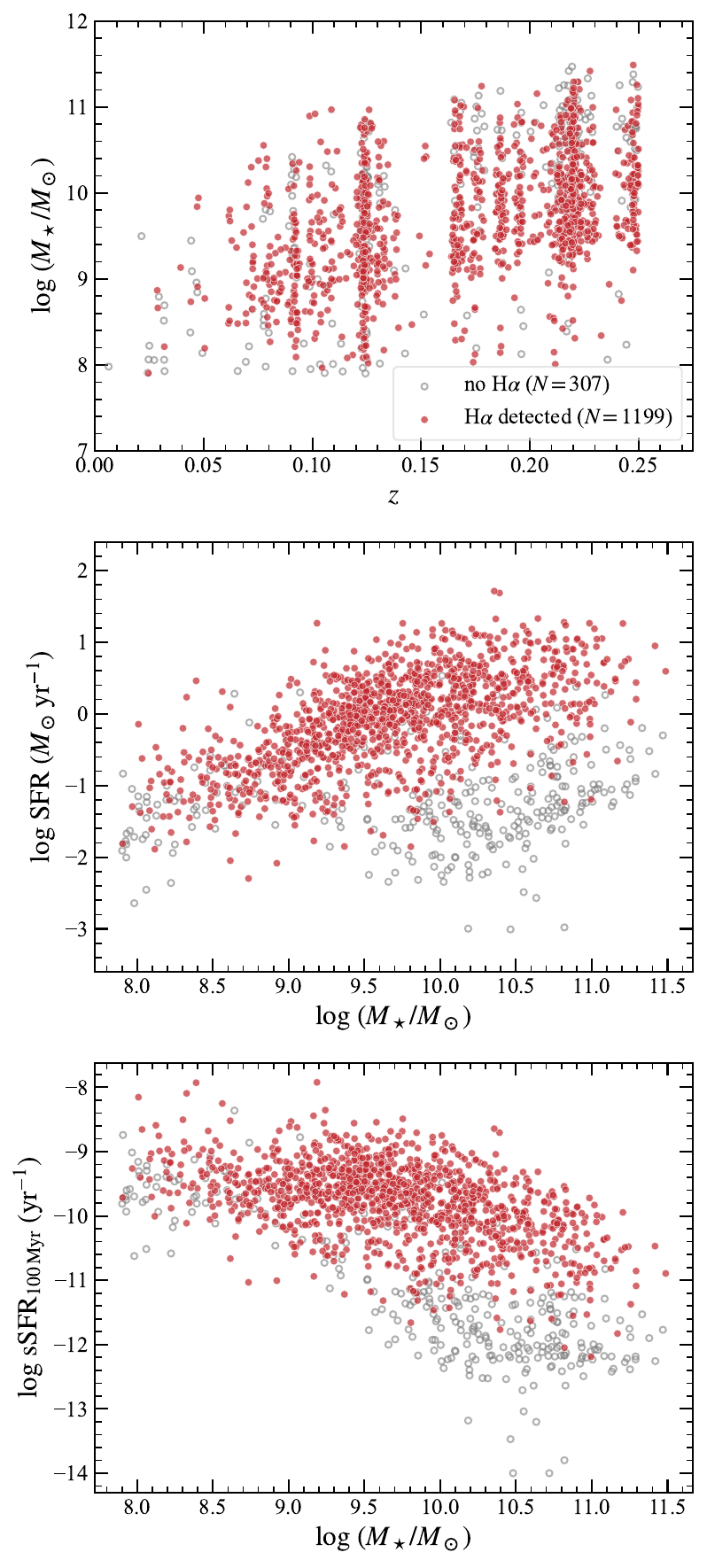}
\caption{Selection diagnostics for our sample. {\em Top}: stellar mass versus redshift. {\em Middle}: SFR versus $M_\star$. {\em Bottom}: sSFR averaged over the past 100 Myr versus $M_\star$. Red solid points show the 1199 galaxies with solid H$\alpha$ detections (S/N $>5$) that comprise our final sample, spanning $7.9<\log M_\star/M_\odot<11.5$. Gray open points lack robust H$\alpha$ detections and are excluded because their ionized-gas velocity dispersions cannot be reliably measured. The SFR and sSFR shown here are obtained from COSMOS2020 Classic catalog.}
\label{fig:selection}
\end{figure}

\begin{figure*}
\centering
\includegraphics[width=\textwidth]{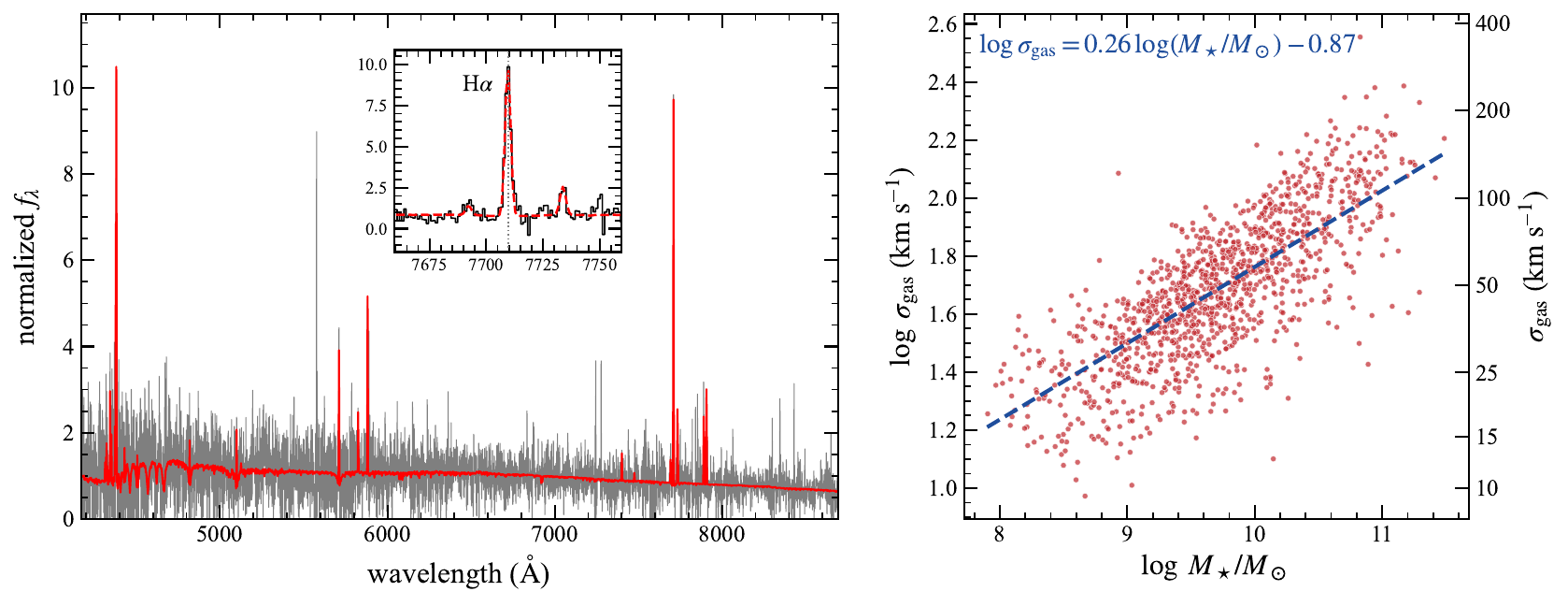}
\caption{{\em Left}: Example full-spectral fitting with {\tt pPXF} for a representative dwarf galaxy ($\log M_\star/M_\odot=8.87$, $z=0.175$), with the H$\alpha$+[N {\sc ii}] region shown in the inset. The observed DESI spectrum is shown in black and the best-fit model in red. {\em Right}: Ionized-gas velocity dispersion $\sigma_{\rm gas}$ versus stellar mass for our full sample, measured from the H$\alpha$ emission line. The blue dashed line shows the ordinary least-squares linear fit, which we adopt as the baseline for computing $\Delta\sigma_{\rm gas}$. \label{fig:masssigma}}
\end{figure*}

DESI is a robotic, fiber-fed multi-object spectrograph on the Mayall 4-m telescope at Kitt Peak National Observatory, deploying 5000 fibers of $1\farcs5$ diameter over a $\sim$8 deg$^2$ field of view, with ten identical three-arm spectrographs covering 3600--9824 \AA\ at a resolution of $R\simeq2000$--$5500$ \citep{DESI2016,DESI2022}, with $R\sim 5000$ at the wavelength of H$\alpha$ emission line for local galaxies. Data Release 1 \citep[DR1;][]{DESI2026} provides spectra and redshifts for more than 18 million unique targets from the first year of main-survey operations.

We cross-match the DESI DR1 spectroscopic catalog with $z<0.25$ with the COSMOS2020 catalog \citep{Weaver2022}, which provides multi-band photometry from UV to near-IR over the COSMOS field. To measure ionized gas kinematics in star-forming galaxies, we require a solid H$\alpha$ emission line detection (${\rm S/N}>5$). AGN and LINER candidates lying above the maximum starburst line of \cite{Kewley2001} are also excluded. This yields a sample of 1199 galaxies with $7.9 < \log M_\star/M_\odot < 11.5$, comprising 459 dwarf galaxies with $7.9<\log M_\star/M_\odot<9.5$, and 740 higher-mass galaxies with $9.5<\log M_\star/M_\odot<11.5$. Compared with the dwarf-only sample of \cite{Luo2026}, our sample extends the stellar mass baseline by more than two orders of magnitude, allowing us to directly search for the mass scale where the kinematic signature associated with feedback-driven outflow cycles disappears. We note that extending the sample to lower stellar masses would substantially reduce the completeness of the DESI data, while the emission lines can be too narrow and/or weak to provide reliable velocity dispersion measurements. Figure \ref{fig:selection} presents the distributions of redshift, stellar mass, SFR, and sSFR for the galaxies in our sample. 

\subsection{Velocity dispersion of ionized gas}
\label{sec:velocity_dispersion}

We measure gas and stellar kinematics simultaneously from each DESI spectrum with the Penalized Pixel-Fitting method ({\tt pPXF}; \citealt{Cappellari2004,Cappellari2017}), which models the stellar continuum with a linear combination of stellar population templates while fitting the emission lines with Gaussian templates. We account for instrumental broadening by convolving the templates with the DESI line-spread function, whose $\sim$1\% uncertainty \citep{Guy2023} has a negligible effect on our results. For the stellar continuum, we adopt the high-resolution ($R\sim10000$) single stellar population models based on the X-shooter Spectral Library \citep{Verro2022}, with the Padova+00 isochrones \citep{Girardi2000} and the \citet{Kroupa2001} initial mass function. To minimize systematic effects from template mismatch, imperfect sky subtraction, and flux-calibration uncertainties, we include a tenth-degree additive Legendre polynomial in the fit. We adopt H$\alpha$ emission line velocity dispersion, $\sigma_{\rm gas}$, as the tracer for further analysis because Balmer emission directly traces recent star formation. The DESI spectral resolution corresponds to a velocity resolution of approximately 25 ${\rm km\ s^{-1}}$ around H$\alpha$ over our redshift range. This is comparable to the lower end of $\sigma_{\rm gas}$ in our sample, indicating that it can be reliably measured. Figure \ref{fig:masssigma} presents an example of our {\tt pPXF} fit. 

Figure \ref{fig:masssigma} shows the correlation between $\sigma_{\rm gas}$ and $M_\star$ for the full sample, consistent with both the results of \cite{Luo2026} and theoretical expectations. Following \cite{Luo2026}, we characterize the correlation with an ordinary least-squares linear fit:
$$\log \sigma_{\rm gas} = (0.26 \pm 0.01) \times \log M_\star/M_\odot - (0.87 \pm 0.06)$$
and define $\Delta \sigma_{\rm gas}$ as the residual relative to this best-fitting relation to remove the mass effect. The uncertainties in the slope and intercept are estimated using bootstrapping. The baseline accounts for the increase in emission line width with the depth of the gravitational potential due to rotation and gravitationally driven turbulence, allowing $\Delta\sigma_{\rm gas}$ to trace excess gas motions potentially associated with feedback-driven outflows (Section~\ref{sec:intro}). Variations in rotation and turbulence at fixed mass can also contribute to $\Delta\sigma_{\rm gas}$, as discussed in Section~\ref{sec:disc_transition}.

\subsection{Spectrophotometric modeling and star formation histories}
\label{sec:prospector}

\begin{figure*}
\centering
\includegraphics[width=1\textwidth]{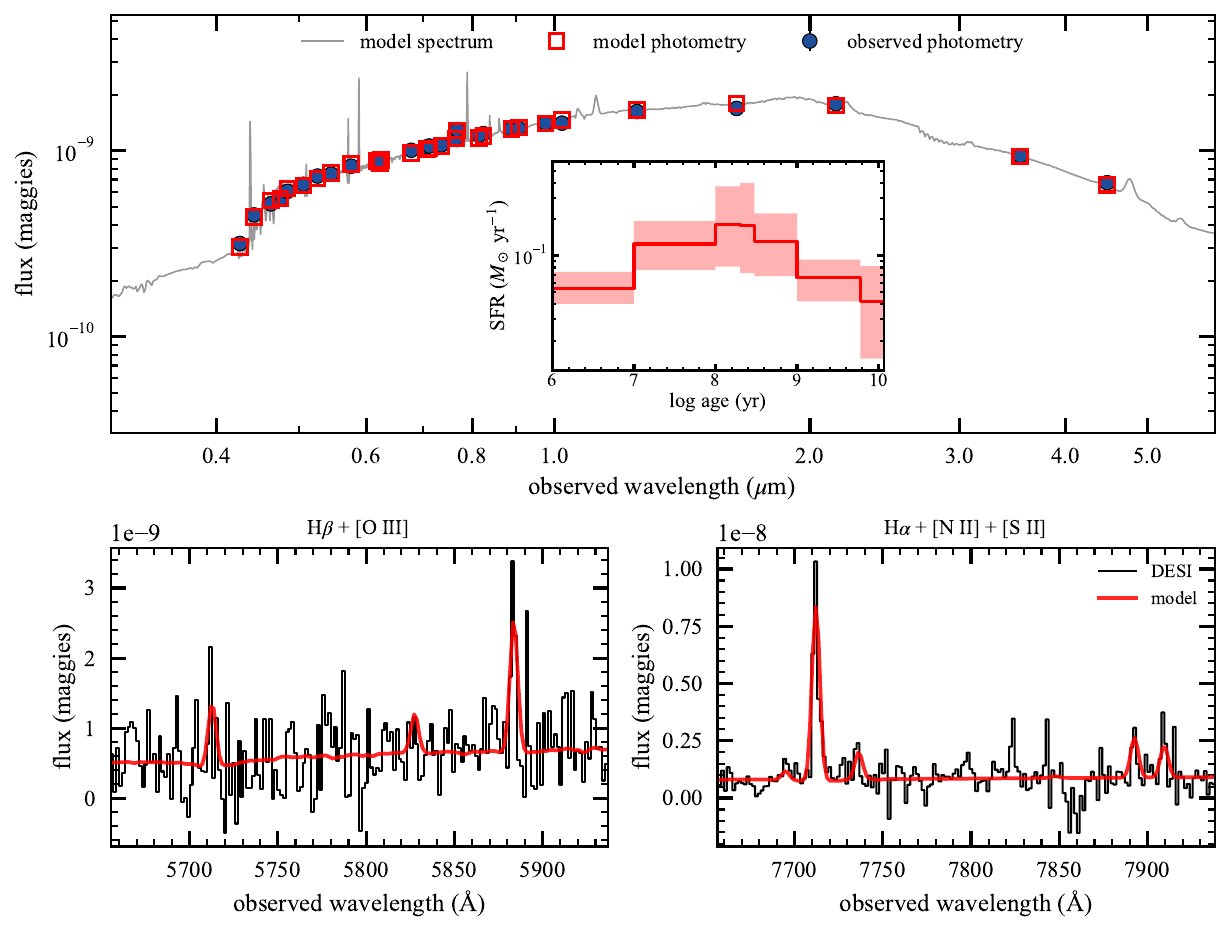}
\caption{Example joint spectrophotometric fit for a representative dwarf galaxy ($\log M_\star/M_\odot=8.87$, $z=0.175$). {\em Top}: Broadband photometry (blue circles), best-fit \texttt{Prospector} model photometry (red squares), and full model spectrum (gray). The inset shows the best-fit SFH, with the shade representing 1-$\sigma$ uncertainty. {\em Bottom}: Zoomed H$\beta$+[\ion{O}{3}] and H$\alpha$+[\ion{N}{2}]+[\ion{S}{2}] regions, comparing the observed DESI spectrum (black) to the best-fit model (red). \label{fig:prospector}}
\end{figure*}

We fit each galaxy jointly to its DESI spectrum and COSMOS2020 photometry using {\tt Prospector} \citep{Johnson2021} with the {\tt FSPS} stellar population synthesis models \citep{Conroy2009}. We adopt the MIST isochrones \citep{Choi2016}, the \cite{Kroupa2001} initial mass function, the \cite{Kriek2013} dust attenuation law, and nebular emission modeled with {\tt CloudyFSPS} \citep{Byler2017}. The photometric data from the COSMOS2020 catalog used in this work include GALEX $NUV$, CFHT/MegaCam $u$, Spitzer/IRAC channels 1 and 2, and the full set of filters from the Subaru and VISTA telescopes listed in \cite{Weaver2022}. We correct the photometry by adding the {\tt LePhare} zero-point offsets for the COSMOS2020 Classic catalog listed in \cite{Weaver2022}, which optimize the absolute calibration in each band. Also, following \cite{Weaver2022}, we add 0.02 mag in quadrature to the optical photometric uncertainties, 0.05 mag to those in $J$, $H$, $K_s$, ch1, and the three narrow bands, and 0.1 mag to that in ch2. For the DESI spectra, we restrict the fit to wavelength regions surrounding the major optical emission lines ([\ion{O}{2}] \(\lambda\lambda3726,3729\), H$\beta$, [\ion{O}{3}] \(\lambda5007\), H$\alpha$, [\ion{N}{2}] \(\lambda6584\), and [\ion{S}{2}] \(\lambda\lambda6717,6731\)), allowing the spectra to primarily constrain the nebular properties and recent star formation while the broadband photometry provides constraints on the overall stellar population. The galaxy redshift is fixed to the value measured from our {\tt pPXF} analysis in Section \ref{sec:velocity_dispersion}.

We reconstruct the SFHs using the non-parametric continuity prior of \cite{Leja2019}, which allows SFR to vary freely between adjacent lookback time bins. We adopt seven age bins spanning [0, 10 Myr], [10 Myr, 100 Myr], [100 Myr, 200 Myr], [200 Myr, 300 Myr], [300 Myr, 1 Gyr], [1 Gyr, 6 Gyr], and [6 Gyr, 14 Gyr]. This directly addresses a limitation noted by \cite{Luo2026}, whose parametric delayed-$\tau$ SFHs yield sSFRs averaged over the past 100 Myr and may therefore smooth out the bursty star formation expected to drive the breathing mode.

We sample the posterior distributions using an implementation of the affine-invariant ensemble sampler \citep{Goodman2010} {\tt emcee} \citep{Foreman-Mackey2013}. Uniform priors are adopted for most free parameters, including the stellar mass ($6 < \log M_\star/M_{\odot} < 12$), stellar metallicity ($-3.5 < {\rm [M/H]} < 0.5$), gas-phase metallicity ($-2.0 < \log Z_{\rm gas}/Z_\odot < 0.5$), ionization parameter ($-4 < \log U < -1$), and dust attenuation ($0 < A_V < 2$ mag for both the birth cloud {\tt dust1} and diffuse dust {\tt dust2} components). The only exception is the SFR ratios between adjacent age bins, for which we adopt the Student's-$t$ prior used in \cite{Leja2019}.

Figure \ref{fig:prospector} presents an example of the joint spectral and photometric fit. From the posterior SFHs, we compute sSFR averaged over the past 10 and 100 Myr, sSFR$_{\rm 10Myr}$ and sSFR$_{\rm 100Myr}$, for further analysis. These two timescales are constrained primarily by H$\alpha$ emission and UV flux, respectively \citep{Kennicutt2012}.

\section{Feedback-driven outflows in low-mass galaxies}
\label{sec:disc_bursty}

\begin{figure}
\centering
\includegraphics[width=\columnwidth]{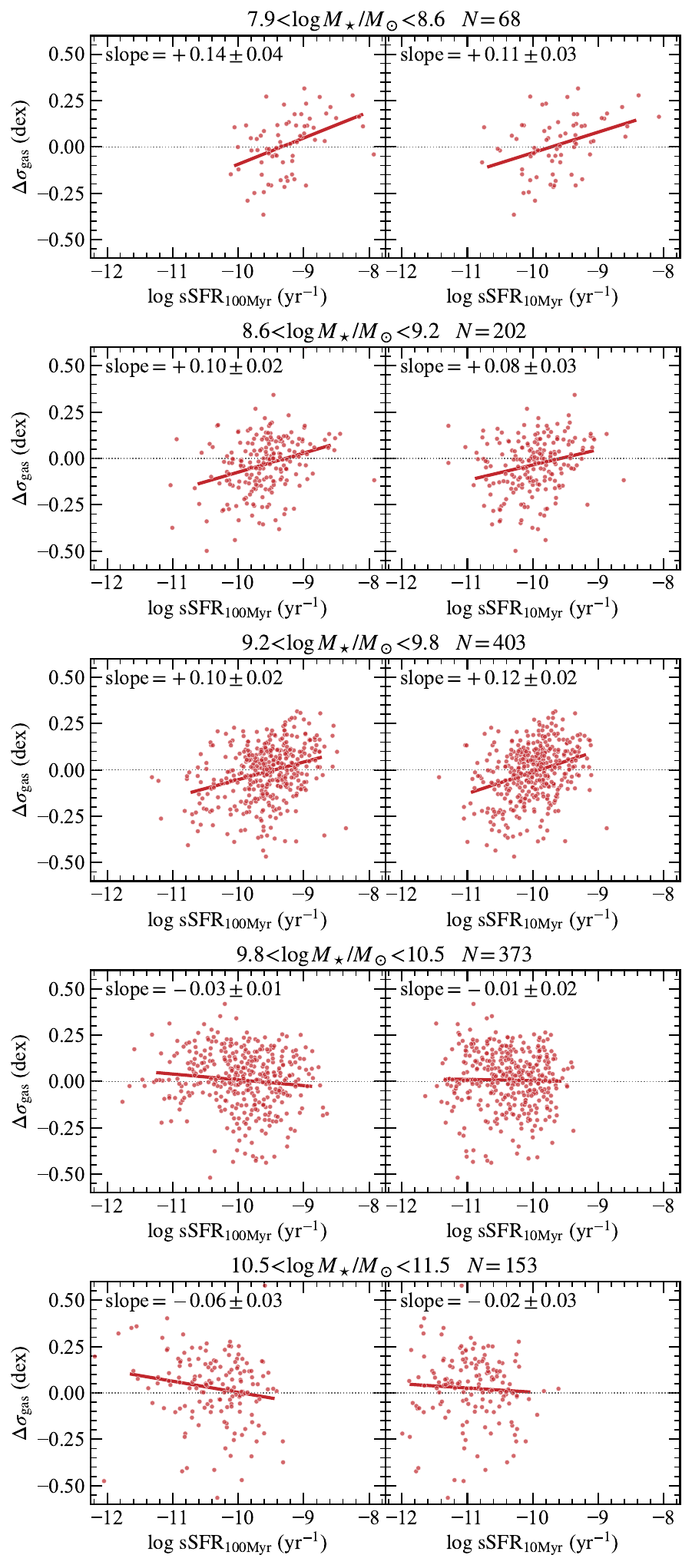}
\caption{Mass-normalized gas velocity dispersion residual, $\Delta\sigma_{\rm gas}$, versus sSFR averaged over 100 Myr (left column) and 10 Myr (right column), in five stellar mass bins spanning $7.9<\log M_\star/M_\odot<11.5$ (rows from top to bottom). Red lines show ordinary least-squares linear fits, with best-fit slopes and sample sizes labeled in each panel. The positive breathing mode signature persists through $9.2<\log M_\star/M_\odot<9.8$ but weakens rapidly above $\log M_\star/M_\odot\sim9.8$. \label{fig:massbins}}
\end{figure}

\begin{figure}
\centering
\includegraphics[width=\columnwidth]{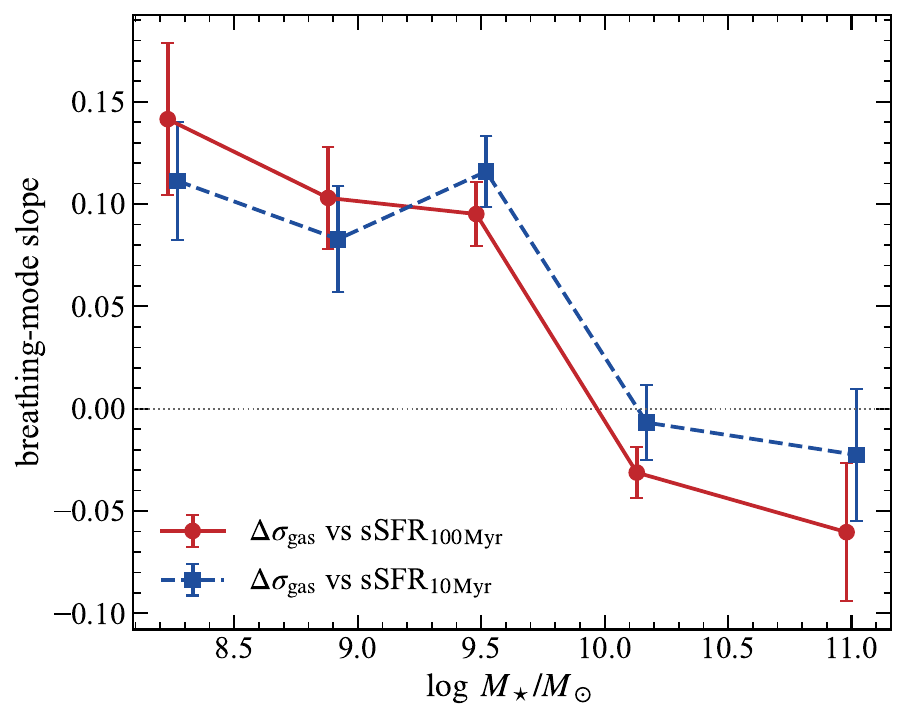}
\caption{Best-fit $\Delta\sigma_{\rm gas}$--sSFR slope from each panel of Figure~\ref{fig:massbins}, plotted against the median stellar mass of its bin, for sSFR$_{\rm 100 Myr}$ (red circles) and sSFR$_{\rm 10 Myr}$ (blue squares). Both indicators show a roughly constant, positive slope of outflow-driven breathing mode at $\log M_\star/M_\odot\lesssim9.8$ that declines through zero above $\log M_\star/M_\odot\sim9.8$. This figure summarizes the central result of this Letter.\label{fig:slope}}
\end{figure}

The central question of this Letter is whether the positive $\Delta\sigma_{\rm gas}$--sSFR relation persists across the full stellar mass range, or whether it is confined to dwarf galaxies where it was originally predicted. To address this question, we divide the sample into five stellar mass bins spanning $7.9<\log M_\star/M_\odot<11.5$ and measure the relation between $\Delta\sigma_{\rm gas}$ and the sSFR averaged over the past 10 and 100 Myr in each bin, shown in Figure \ref{fig:massbins}. The bins are chosen to contain comparable numbers of galaxies while resolving the dwarf regime, where the signature is expected to be strongest. Slope uncertainties are estimated from bootstrap resamplings of the full analysis, including refitting the $\sigma_{\rm gas}$--$M_\star$ relation.

Throughout the low-mass regime ($\log M_\star/M_\odot\lesssim9.8$), $\Delta\sigma_{\rm gas}$ correlates positively with both sSFR indicators in all three stellar mass bins. The best-fitting slopes range from 0.10 to 0.14 for sSFR$_{\rm 100Myr}$ and from 0.08 to 0.12 for sSFR$_{\rm 10Myr}$, consistent with the FIRE-2 predictions ($0.092 \pm 0.005$ for sSFR$_{\rm 100Myr}$ and $0.115 \pm 0.004$ for sSFR$_{\rm 10Myr}$) and the measurements from observation ($0.065 \pm 0.012$ for sSFR$_{\rm 100Myr}$ and $0.096 \pm 0.012$ for sSFR$_{\rm 10Myr}$) of \cite{Luo2026}. Our lowest-mass bin has a somewhat steeper slope than the two adjacent dwarf bins, but it is also the smallest subsample and the difference is within the uncertainties.

Both sSFR averaging windows give similar slopes, suggesting that the signal is not tied exclusively to either the youngest ($<10\ {\rm Myr}$) stellar populations or longer-term ($100\ {\rm Myr}$) star formation activity. This is broadly consistent with breathing cycles lasting a few hundred Myr in FIRE-2 \citep{ElBadry2016,ElBadry2017}. Overall, these results support the interpretation of the positive $\Delta\sigma_{\rm gas}$--sSFR correlation as a kinematic signature of feedback-driven outflows.

\section{A transition near $\log M_\star/M_\odot\sim 9.8$}
\label{sec:disc_transition}

In Figure \ref{fig:massbins}, we observe that the positive $\Delta\sigma_{\rm gas}$--sSFR correlation does not extend to higher stellar masses. Above $\log M_\star/M_\odot \sim 9.8$, the positive correlation rapidly weakens. In the $9.8<\log M_\star/M_\odot<10.5$ bin, the relation becomes close to zero for both sSFR indicators. At higher masses ($10.5<\log M_\star/M_\odot<11.5$), the slope measured using sSFR$_{\rm 100Myr}$ seems to become negative, whereas the sSFR$_{\rm 10Myr}$ slope remains more consistent with zero. Figure \ref{fig:slope} summarizes this mass dependence and shows that both indicators trace a rapid decline in the correlation from the dwarf regime, placing the transition around $\log M_\star/M_\odot\sim 9.5$--10. The transition is slightly more pronounced for sSFR$_{\rm 100Myr}$ compared to sSFR$_{\rm 10Myr}$. This rapid mass dependence is a principal new result of this Letter.

The stellar mass at which the signature disappears can be directly compared with the range over which the breathing mode operates in the simulations that predicted it. \cite{ElBadry2016} found that feedback-driven fluctuations are strongest at $M_\star \sim 10^{7}$--$10^{9.6}\ M_\odot$ in FIRE-2 galaxies, the same regime in which feedback most efficiently forms DM cores. Outside this window, the effect weakens because feedback is either too weak to perturb the potential in the lowest-mass galaxies or suppressed by deeper potential wells and lower gas fractions in the most massive galaxies. The upper edge of the FIRE-2 window therefore lies between our third and fourth stellar mass bins, with the positive correlation detected throughout $9.2<\log M_\star/M_\odot<9.8$ but absent in $9.8<\log M_\star/M_\odot<10.5$. Given our bin widths, the observed transition is consistent with the $\log M_\star/M_\odot \sim 9.6$ upper limit predicted by \cite{ElBadry2016}, providing the first observational test of this prediction.

The physical origin of a characteristic mass in this range was anticipated long before the simulations. \cite{DekelSilk1986} argued that supernova-driven winds can expel a substantial fraction of the gas only from haloes with virial velocities $V\lesssim100\ {\rm km\ s^{-1}}$, where supernova energy becomes comparable to the gas binding energy. \cite{DekelWoo2003} showed that this energy balance sets a ``fundamental line'' that terminates near $V \sim 100\ {\rm km\ s^{-1}}$, above which supernova feedback can no longer regulate star formation \citep[see also][]{Dekel2019}. This threshold corresponds to $M_{\rm halo} \sim 3\times10^{11}\ M_\odot$ and $M_\star \sim 10^{10}\ M_\odot$ today \citep{Moster2013}, close to our observed transition mass. Although this analytic threshold and the FIRE-2 window share the same underlying energy criterion and are not independent predictions, their agreement supports an observational interpretation of the characteristic mass as the supernova-feedback threshold.

A decline in the kinematic signature of outflow-driven breathing mode is therefore expected as the potential well deepens and the gas fraction within the effective radius falls \citep{ElBadry2016}. More massive systems also have lower wind mass loading and increasingly retain and recycle gas rather than driving strongly ejective cycles \citep{Muratov2015,Christensen2016}. At the same time, gravitational instability, accretion, and radial transport can make larger contributions to gas motions in high-dispersion disks \citep{KrumholzBurkhart2016,Krumholz2018}.

The subtle difference between the two sSFR timescales may provide an additional clue. Above the transition, sSFR$_{\rm 100Myr}$ may trace longer-lived connections among gas supply, morphology, and dynamical support. At fixed mass, extended rotation-dominated systems could have higher sSFR and narrower integrated lines than centrally concentrated systems. A short-timescale feedback response could partially offset this structural trend in sSFR$_{\rm 10Myr}$, resulting in the flatter relation. However, the difference could simply arise from larger uncertainties in the youngest SFH bin.

A complementary perspective comes from medium-band imaging surveys such as Merian \citep{Luo2024,Danieli2025}, which isolate emission line flux from the stellar continuum to provide spatially resolved ionized gas maps for large samples. Using Merian H$\alpha$ maps, \citet{Mintz2024} found that high sSFR, low-mass galaxies exhibit compact, heterogeneous H$\alpha$ emission, whereas less active systems show more diffuse emission. Although they did not constrain a transition mass, this trend was driven primarily by galaxies with $\log M_\star/M_\odot<9.3$. Combining such maps with ionized gas kinematics could test whether concentrated star formation is associated with enhanced velocity dispersion and whether this connection weakens at higher mass.

Several observational systematics could in principle contribute to the observed mass dependence. First, residual AGN or LINER-like emission may broaden the emission lines, particularly in massive, low-sSFR galaxies, although our sample excludes galaxies above the \cite{Kewley2001} maximum starburst line. Second, redshift-dependent effects may arise because the physical region covered by the DESI fiber increases with redshift, enhancing the contribution from unresolved rotation and aperture coverage \citep{DESI2022,Varidel2020}, while the stellar mass completeness of our sample also decreases. Finally, the H$\alpha$ selection preferentially excludes low-sSFR galaxies, potentially introducing a mass-dependent selection bias. We test these possibilities with a stricter emission-line classification, a restricted redshift range, and an sSFR cut, respectively (Appendix \ref{sec:appendix_1}). The overall mass dependence remains qualitatively unchanged in all these tests.

\section{Implications for the breathing mode beyond gas kinematics}
\label{sec:disc_cores}

In the hydrodynamic simulations that motivated this work, the breathing mode is a chain of linked predictions. Bursty star formation drives gas outflows, while the outflows and subsequent re-accretion produce fluctuations in the central potential. These fluctuations can heat stellar orbits, drive radial migration, and cause the effective radius to vary by factors of 2--3 on $\sim200$ Myr timescales. Integrated over many cycles, they can also transform central DM cusps into cores \citep{Governato2010,Pontzen2012,ElBadry2016,ElBadry2017}. The positive $\Delta\sigma_{\rm gas}$--sSFR correlation below $\log M_\star/M_\odot\sim9.8$ supports the first link in this chain, showing that recent star formation and disturbed ionized-gas motions vary together with slopes matching the FIRE-2 predictions. The disappearance of this correlation near the mass where supernova feedback is expected to lose its grip on the gas further supports this picture. However, the subsequent predictions have not yet been reproduced consistently in observations.

The predicted fluctuations of the stellar component provide the clearest comparison. \cite{Patel2018} tested the predicted anti-correlation between SFR and half-light radius in isolated dwarfs but found the opposite trend, with higher-sSFR galaxies being more extended and disk-dominated. At lower masses, \cite{Emami2021} found only weak or absent correlations between stellar size and recent star formation, consistent with FIRE-2 predictions but with limited discriminating power because the stellar size responds to bursts with a $\sim50$ Myr delay. Stellar morphology shows a similar tension. Observed dwarfs are predominantly thick-disk-like in their inner regions and often host rotation-supported gas disks \citep{ElBadry2018b,KadoFong2020}, whereas simulations produce a substantial population of dispersion-supported, ``non-disky'' dwarfs with burstier recent star formation \citep{ElBadry2018b,KadoFong2022,Zeng2024,Celiz2025}.

The mismatch may reflect both differences in stellar response and differences between the simulated and observed environments. Once formed, stellar disks may be harder to disrupt than gas, requiring sufficiently strong and frequent potential fluctuations to heat the collisionless component. A feedback model calibrated to reproduce bursty gas behavior could therefore still overpredict stellar dynamical heating. \cite{Dekel2020} argue that gas disks survive primarily in haloes above $M_{\rm halo}\sim2\times10^{11}\ M_\odot$, corresponding to $M_\star\sim10^{9}\ M_\odot$ \citep{Moster2013}, placing the observed disky dwarfs at $10^{8.5}$--$10^{9.5}\ M_\odot$ near the regime where feedback and disk survival are both sensitive to the details of the feedback model. At the same time, most high-resolution simulations of dwarfs follow isolated field systems, whereas an observed sample such as ours spans a broad range of environments. If disks form more readily under particular conditions (e.g., through coherent accretion along cosmic filaments), or if dwarfs in groups and clusters are additionally shaped by tidal interactions and ram pressure \citep[e.g.,][]{Geha2012,Jiang2019}, then observations would naturally show greater morphological diversity.

Whether the same feedback cycles produce the predicted changes in stellar size, morphology, and kinematics, and ultimately transform DM cusps into cores, remains less settled. The mass dependence of the $\Delta\sigma_{\rm gas}$--sSFR slopes provides a useful benchmark for testing this part of the model. Future work should combine gas kinematics with stellar sizes, intrinsic shapes, and rotational support for star-forming galaxies, while extending the comparison to dwarfs in different environments. Such measurements will help determine whether the observed gas-kinematic response is accompanied by the stellar and dynamical changes predicted by the breathing mode scenario, and whether environmental effects contribute to the remaining tension between observations and simulations.

\section{Summary}
\label{sec:summary}

Using $\sim$1200 star-forming galaxies with DESI spectroscopy and COSMOS2020 photometry spanning $7.9<\log M_\star/M_\odot<11.5$ within $z<0.25$, we perform the first systematic observational test of the stellar feedback-driven ``breathing mode'' across the transition from dwarf to massive galaxy regimes. DESI’s high spectral resolution at H$\alpha$ ($R \sim 5000$) enables reliable measurements of the gas velocity dispersions of low-mass galaxies. To capture bursty recent star formation, we jointly fit the DESI spectra and COSMOS2020 photometry with Prospector, reconstructing the non-parametric SFHs and inferring sSFR averaged over the past 10 and 100 Myr. Our main results are summarized as follows:

\begin{itemize}
\item With an independent spectroscopic sample and non-parametric SFH reconstruction, we confirm the positive correlation between mass-normalized gas velocity dispersion $\Delta\sigma_{\rm gas}$ and sSFR previously reported by \cite{Luo2026} throughout the low-mass galaxy regime ($\log M_\star/M_\odot \lesssim 9.8$), obtaining slopes of 0.08-0.14 that are in close agreement with the FIRE-2 predictions. 

\item The correlation weakens rapidly above $\log M_\star/M_\odot \sim 9.8$ and is consistent with zero at higher masses, providing the first empirical constraint on the transition stellar mass where the kinematic signature of the breathing mode becomes negligible.

\item The disappearance of the positive correlation is consistent with the $\log M_\star/M_\odot \sim 9.6$ upper limit of the breathing mode window predicted by FIRE-2 and with the $\log M_\star/M_\odot \sim 10$ threshold above which the effect of supernova feedback is expected to decline. 

\end{itemize}

Overall, our results suggest that the observable kinematic response to stellar feedback becomes subdominant above a characteristic stellar mass of $M_\star\sim10^{10}\ M_\odot$, as hydrodynamical simulations predict. This transition provides an empirical benchmark for models of feedback regulated galaxy evolution and the processes linked to dark matter core formation, and motivates both cosmological simulations and spatially resolved spectroscopic observations spanning this mass range. A crucial next step is to combine this gas kinematic signature with stellar structural properties, such as galaxy morphology or stellar kinematics, to test the feedback-driven breathing mode scenario more comprehensively. Current and forthcoming deep and wide imaging surveys such as Euclid, Rubin, and Roman will provide unprecedented constraints on the morphologies and structures of low-mass galaxies, while spectroscopy from DESI, 4MOST, and Spec-S5 will provide the complementary kinematic and stellar population diagnostics needed to extend this test to much larger samples.

\begin{acknowledgments}

We thank the referee for useful feedback. This research used data obtained with the Dark Energy Spectroscopic Instrument (DESI). DESI construction and operations is managed by the Lawrence Berkeley National Laboratory. This material is based upon work supported by the U.S. Department of Energy, Office of Science, Office of High-Energy Physics, under Contract No. DE–AC02–05CH11231, and by the National Energy Research Scientific Computing Center, a DOE Office of Science User Facility under the same contract. Additional support for DESI was provided by the U.S. National Science Foundation (NSF), Division of Astronomical Sciences under Contract No. AST-0950945 to the NSF’s National Optical-Infrared Astronomy Research Laboratory; the Science and Technology Facilities Council of the United Kingdom; the Gordon and Betty Moore Foundation; the Heising-Simons Foundation; the French Alternative Energies and Atomic Energy Commission (CEA); the National Council of Humanities, Science and Technology of Mexico (CONAHCYT); the Ministry of Science and Innovation of Spain (MICINN), and by the DESI Member Institutions: www.desi.lbl.gov/collaborating-institutions. The DESI collaboration is honored to be permitted to conduct scientific research on I’oligam Du’ag (Kitt Peak), a mountain with particular significance to the Tohono O’odham Nation. Any opinions, findings, and conclusions or recommendations expressed in this material are those of the author(s) and do not necessarily reflect the views of the U.S. National Science Foundation, the U.S. Department of Energy, or any of the listed funding agencies.

\end{acknowledgments}

\begin{contribution}

Yimeng Tang, Yifei Luo, and Conghao Zhou contributed equally to this work.

\end{contribution}

\restartappendixnumbering
\appendix

\section{Robustness of the Mass-dependent $\Delta\sigma_{\rm gas}$--sSFR Relation}
\label{sec:appendix_1}

Section \ref{sec:disc_transition} shows that the $\Delta\sigma_{\rm gas}$--sSFR correlation weakens rapidly above $\log M_\star/M_\odot\sim9.8$. Here we test whether this mass dependence could arise from several observational effects by repeating the analysis with alternative sample selections respectively. The fitting procedure and stellar mass bins are unchanged from those used in Figures \ref{fig:massbins} and \ref{fig:slope}.

\begin{figure}
\centering
\includegraphics[width=\columnwidth]{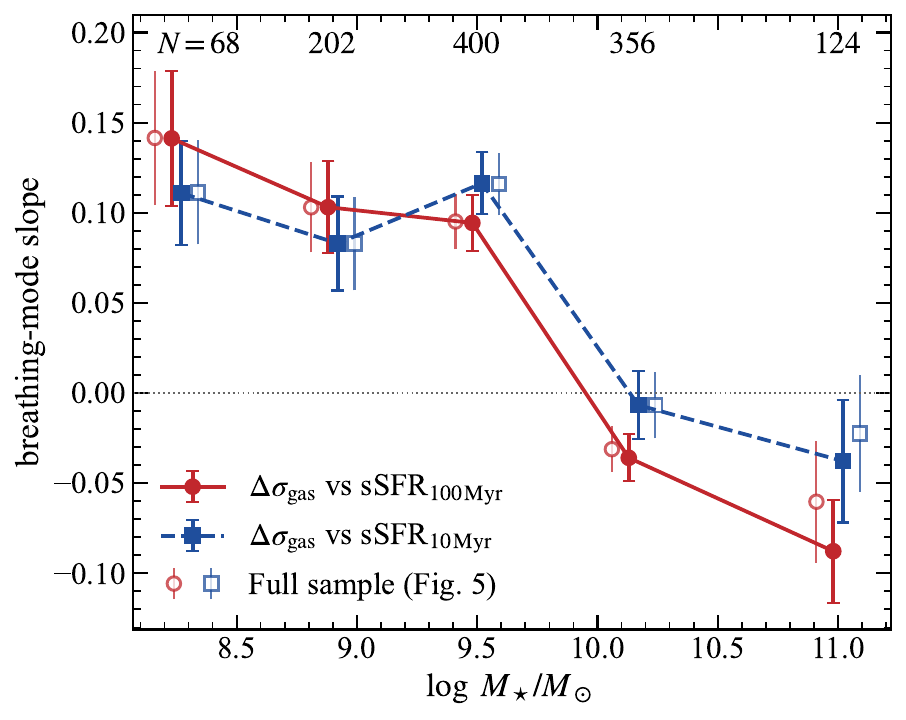}
\caption{Same as Figure \ref{fig:slope}, but only galaxies below \cite{Kauffmann2003} are included. The numbers at the top indicate the number of galaxies in each stellar mass bin, and the open symbols show the full-sample slopes from Figure~\ref{fig:slope}.}
\label{fig:slope_kau03cut}
\end{figure}

First, residual AGN or LINER-like emission could preferentially affect massive, low-sSFR galaxies and broaden their emission lines. Our fiducial sample excludes galaxies above the \cite{Kewley2001} maximum starburst line. We here adopt a more conservative selection by retaining only galaxies below the \cite{Kauffmann2003} line, which selects a purer star-forming population. The results are shown in Figure \ref{fig:slope_kau03cut}.

\begin{figure}
\centering
\includegraphics[width=\columnwidth]{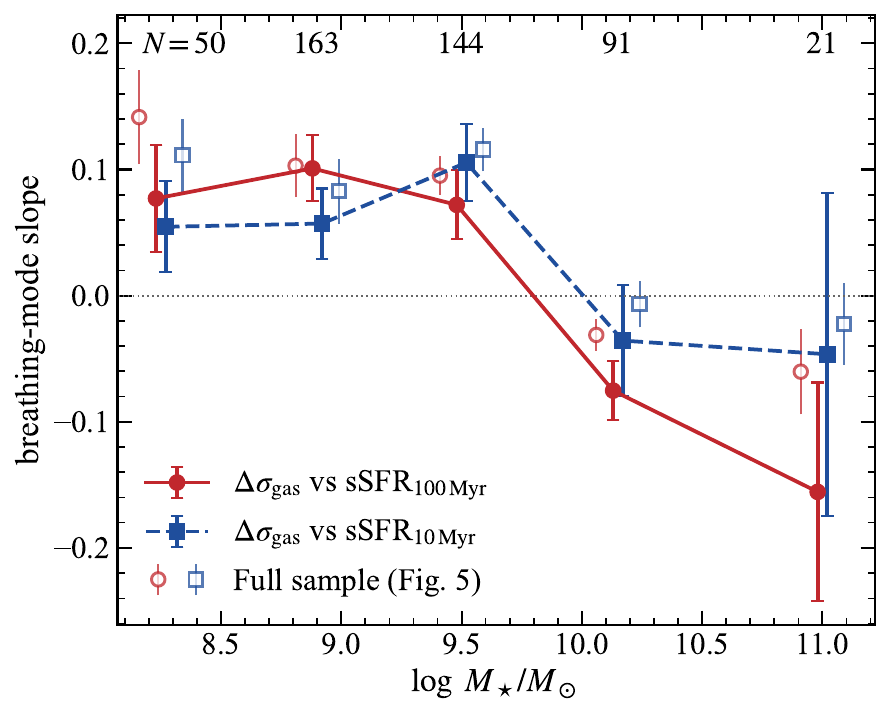}
\caption{Same as Figure \ref{fig:slope}, but only galaxies with $z<0.15$ are included. The numbers at the top indicate the number of galaxies in each stellar mass bin, and the open symbols show the full-sample slopes from Figure~\ref{fig:slope}.}
\label{fig:slope_zcut}
\end{figure}

Second, the physical region covered by the DESI fiber increases with redshift, sampling approximately 0.6 kpc at $z=0.02$ but 5.9 kpc at $z=0.25$. Unresolved rotation and aperture coverage could therefore introduce mass- or sSFR-dependent trends. The stellar mass completeness of our sample also declines toward higher redshift. To reduce both effects, we repeat the analysis using only galaxies at $z<0.15$, where the physical aperture range is narrower and the sample is more nearly complete in stellar mass (Figure \ref{fig:slope_zcut}).

\begin{figure}
\centering
\includegraphics[width=\columnwidth]{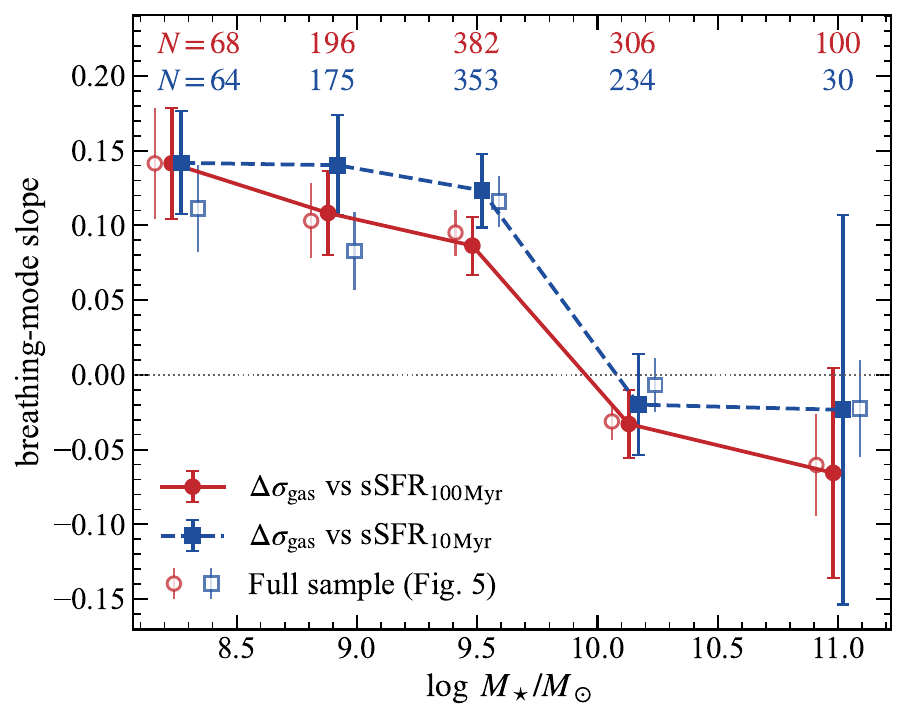}
\caption{Same as Figure \ref{fig:slope}, but only galaxies with ${\rm sSFR}>-10.5$ are included. The red and blue numbers at the top indicate the number of galaxies in each stellar mass bin used for the sSFR$_{\rm 100Myr}$ and sSFR$_{\rm 10Myr}$ relations, respectively. The open symbols show the full-sample slopes from Figure~\ref{fig:slope}.}
\label{fig:slope_ssfrcut}
\end{figure}

Finally, the H$\alpha$ S/N requirement preferentially removes low-sSFR galaxies, with a stronger effect at higher stellar mass. This mass-dependent selection bias could affect the measured slopes if the low-sSFR end is sampled differently across the stellar-mass bins. To test this, we follow \cite{Luo2026} and restrict the sample to $\log{\rm sSFR}>-10.5$, with the results shown in Figure \ref{fig:slope_ssfrcut}.

In all these cases, the stellar mass dependence remains qualitatively consistent with our results in Figure~\ref{fig:slope}. The restricted samples produce larger uncertainties in some of the mass bins because of the reduced sample size and sSFR dynamic range. These tests indicate that the observed transition is not primarily driven by residual AGN or LINER contamination, redshift-dependent aperture effects, or H$\alpha$ selection.

\bibliography{reference}{}
\bibliographystyle{aasjournalv7}

\end{document}